\documentstyle[aps,prl,floats,epsfig]{revtex}

\newcommand{\bastar}{\begin{eqnarray*}}
\newcommand{\eastar}{\end{eqnarray*}}
\newskip\humongous \humongous=0pt plus 1000pt minus 1000pt

\newif\ifdtup

\relax
\newcommand{\be}{\begin{equation}}
\newcommand{\ee}{\end{equation}}
\newcommand{\bea}{\begin{eqnarray}}
\newcommand{\eea}{\end{eqnarray}}

\newcommand{\tF}{{\tilde F}}

\newcommand{\dfrac}{\displaystyle\frac}
\newcommand{\ba}{\begin{array}}
\newcommand{\ea}{\end{array}}

\newcommand{\nn}{\nonumber}

\newcommand{\x}{{\tilde x}}
\newcommand{\y}{{\tilde y}}
\begin{document}
%\twocolumn[\hsize\textwidth\columnwidth\hsize\csname@twocolumnfalse%
%\endcsname
\title  {Reply to the Comment by U. Jentschura and E. Weniger}
\bigskip
\author{
D. G. Pak}

\address{
Asia Pacific Center for 
Theoretical Physics, 207-43 Cheongryangri-dong, Dongdaemun-gu,
               Seoul 130-012 Korea\\
{\scriptsize
dmipak@mail.apctp.org} \\ \vskip 0.3cm
}
\maketitle
\begin{abstract}
It is shown that the criticism (revised version) 
made in hep-th/0007108 by Jentschura and
Weniger on hep-th/0006057 has no valid ground. 
Furthermore we emphasize that the concept of the electric-magnetic duality
used in the analysis of QED one-loop effective action in physics/0010038
has first been introduced in hep-th/0006057.

\vspace{0.3cm}
\end{abstract}

%\narrowtext
%\widetext
\bigskip
%                           ]

Recently we have presented a non-perturbative but convergent 
series representation of the one-loop effective action
in the standard and scalar QED, and established the existence of  
a new electric-magnetic duality in these gauge theories 
at the quantum level \cite{cho1}.
But subsequently Jentschura and Weniger \cite{bugs}
claimed that ``the results in \cite{cho1} have appeared 
in the literature before'', in particular in \cite{miel}. 
In the first version of the present note I showed that
their criticism is not based on the facts. But they made the same false
assertion in their revised Comment. Here I refute their assertion
item by item. 

\section{Sitaramachandrarao's Identity}

First, Jentschura and Weniger claimed (in their first version of \cite{bugs})
that ``the Sitaramachandrarao's
identity (6) in \cite{cho1},
which plays a crucial role in the derivation of the effective action of QED,
corresponds to the identity (2.8) in \cite{miel}''. It is surprising how
they could make such a false claim without checking the elementary 
facts. Obviously they are different, although superficially
they look similar. More importantly {\it the identity
(2.8) in \cite{miel} is simply incorrect}. This can easily be seen  
(to those who have read \cite{miel}) even in 
the simple case of pure magnetic background.
Now, Jentschura and Weniger in their revised Comment \cite{bugs} assert
that the identity (2.8) in \cite{miel} had ``typographical errors'', and 
insist that it becomes identical to the Sitaramachandrarao's 
identity ``after correcting these typographical errors''.
To see whether this is true or not, we show here both the Sitaramachandrarao's
identity \cite{cho1,sita} 
\bea
& xy \coth x \cot y = 1 + \dfrac{1}{3} (x^2-y^2)  \nn \\
& -  \dfrac{2}{\pi}  x^3 y  \sum_{n=1}^{\infty} \dfrac{1}{n} \dfrac{\coth
(\dfrac{n \pi y}{x})}{(x^2 +n^2  \pi^2)}   +\dfrac{2}{\pi}    x y^3
\sum_{n=1}^{\infty} \dfrac{1}{n}
 \dfrac{\coth (\dfrac{n \pi  x}{y})}{(y^2 - n^2 \pi^2)},
\eea
and the Mielniczuk's identity (2.8) in \cite{miel}
\bea
& \x \y u^2 \coth (\x u) \cot (\y u) -1 - \dfrac{1}{3} (\x^2-\y^2) u^2  \nn \\
& = - \dfrac{2 \x \y^3 u^4}{\pi} \sum_{k=1}^{\infty} \dfrac{1}{k}
\dfrac{1}{(\y^2 u^2+k^2\pi^2)}\coth{\Big (}\dfrac{\x}{\y} k\pi{\Big )} \nn\\
&+ \dfrac{2 \x \y^3 u^4}{\pi} \sum_{k=1}^{\infty} \dfrac{1}{k}
\dfrac{1}{(\y^2 u^2 - k^2 \pi^2)} 
\coth {\Big (}\dfrac{\x}{\y} k \pi{\Big )} \nn \\
& + {\rm the~~ expression~~ obtained~~ from~~ the~~ first~~
component~~ by~~ the~~ replacement~~ rule } \nn \\
& \x \rightarrow - i \y, ~~~~~ \y \rightarrow -i \x . 
\eea
Evidently they are different. Now, Jentschura and Weniger simply 
manipulate and rewrite the Mielniczuk's identity (2) at their will, and claims
that the two identities becomes identical after the ``correction
of the typographical errors'' in (2). 
They claim this by asserting that ``the second term 
on the right-hand side of the identity (2)
should be ignored'', and furthermore ``$\x$ and $\y$ should be interchanged
in the first term on the right-hand side of (2)''. And they insist this by
``{\it assuming} that Mielniczuk {\it may} have checked the 
consistency'' of his identity (2). {\it We} agree that he must
have checked his identity, so that one should take his expression (2) as it is.
More importantly we believe that ``ignoring'' 
an essential part of the equation is {\it not} a
`` correction of the typographical errors'', but an intentional falsification 
to make a wrong equation correct.  Notice that they 
could have ``corrected the typographic errors'' of (2) in a totally
different way, for example, simply by ignoring the first term 
on the right-hand side of (2). So it is puzzling why they insist
to ignore the second term, rather than the first term, on the right-hand side
of (2). Obviously this type of manipulation is not ``a correction
of typographic errors''. 

In fact, whether one should regard this kind of manipulation
as an acceptable ``correction of the typographic errors'' or not may not
even be the central issue. {\it The central issue here is that,
without knowing the correct Sitaramachandrarao's identity,
how one could possibly ``correct the typographic errors'' in (2)}. 
This is the central issue. Without (1), could Jentschura and Weniger
still insist that `` the second term.....should be ignored'',
and ``$\x$ and $\y$ should be interchanged.....''? On what ground?
Under this circumstance one can not stop wondering how they
would have ``corrected the typographic errors'' of 
the Mielniczuk's identity, {\it had they
not known in advance} the Sitaramachandrarao's identity
from \cite{cho1}. Had the correct identity assumed a different
form, they probably would have insisted a different ``correction 
of the typographic errors'' to match (2) with (1).  

Notice that when Mielniczuk (with other collaborators) reviewed and 
discussed the same subject eleven years later \cite{m2}, {\it he 
himself has never mentioned about or pointed out the possible 
``typographical errors'' of his previous work \cite{miel}}. 
As Jentschura and Weniger have assumed, he must have checked 
the consistency of his equation (2). Had he found
any ``typographic error'', he should have corrected them in
\cite{m2}. He didn't. To us this is another 
indication which suggests that Jentschura and Weniger are falsifying 
the Mielniczuk's identity at their will. 

We emphasize again that the Sitaramachandrarao's identity is 
a rigorous identity which is non-trivial even from 
the mathematical point of view. 
A preliminary version of the identity which was expressed
as a divergent asymptotic series was first obtained by
Ramanujan, and only later 
Sitaramachandrarao has improved the Ramanujan's identity 
and obtained the above convergent expression. 
Even to Ramanujan the identity (1) was not obvious!
Moreover this identity had been known only among very
few mathematicians in India until recently, when the Ramanujan's
Notebooks was printed and circulated in 1989 \cite{sita}.
This implies that it would be very difficult to ``correct the typographic
errors'' of (2) without knowing the correct Sitaramachandrarao's
identity.

Of course, the final judgement on whether the Mielniczuk's identity (2)
is indeed ``correct'' but has ``typographic errors'' should ultimately come
from the audience, in particular from the experts working in this field.
The Mielniczuk's paper has been around more than eighteen years,
and after the work \cite{miel} numerous papers and review articles \cite {p1,p2}
have been published on this subject.
But as far as we understand, {\it no single independent
expert has appreciated his work so far} \cite{p1,p2}. 
This means that the experts in this field evidently
knew about the critical defect in \cite{miel}, and 
did not take it seriously.  Curiously, only after our correct
result has been publicized Jentschura and Weniger start to insist that
one should give the credit to \cite{miel}. It is really puzzling why they
didn't insist this earlier.

\section {Effective Action and Renormalization}

Secondly, our effective action is clearly different 
from the expression in \cite{miel}.
In fact using the wrong identity (2.8) one can not possibly derive a correct
effective action. {\it An honest derivation 
of the effective action using the correct identity should 
produce both the real and the imaginary
parts simultaneously}, as we showed in \cite{cho1}.  To see this we start 
from the well-known integral expression
\bea
\Delta {\cal L} = - \dfrac{1}{8\pi^2} ab \int_{0+i\epsilon}
^{\infty+i\epsilon} \dfrac{dt}{t}
\coth (at) \cot (bt) e^{-m^2 t},
\eea
where
\bea
a = \dfrac{e}{2} \sqrt {\sqrt {F^4 + (F \tF)^2} + F^2}, 
~~~~~b = \dfrac{e}{2} \sqrt {\sqrt {F^4 + (F \tF)^2} - F^2}. \nn
\eea
Notice that here the above contour of the integral is dictated by
the causality. Now, carrying out the integral with 
the Sitaramachandrarao's identity (1), we automatically obtain
the following effective action \cite{cho1}
\bea
&{\cal L}_{eff} = -\dfrac{a^2 - b^2}{2 e^2}(1-\dfrac{e^2}{12\pi^2} 
\ln\dfrac{m^2}{\mu^2}) \nn\\
&- \dfrac{1}{4\pi^3} ab  \sum_{n=1}^{\infty}\dfrac{1}{n}
{\Big [}\coth(\dfrac{n \pi b}{a}){\Big (} {\rm ci}(\dfrac{n \pi m^2}{a}) 
\cos(\dfrac{n \pi m^2}{a}) 
+{\rm si}(\dfrac{n \pi m^2}{a}) \sin(\dfrac{n \pi m^2}{a}){\Big )}\nn\\
&- \dfrac{1}{2} \coth ( \dfrac{n \pi a}{b}){\Big (} \exp(\dfrac{n \pi m^2}{b})
{\rm Ei}(-\dfrac{n \pi m^2}{b} )   
+ \exp(-\dfrac{n \pi m^2}{b} ) {\rm Ei}(\dfrac{n \pi m^2}{b} 
- i \epsilon){\Big )}{\Big ]}.
\eea
In contrast
the derivation in \cite{miel} with (2.8) could not produce any information
on the imaginary part of the action whatsoever.  
So it is obvious that our result is different
from \cite{miel}. 

Later Mielniczuk with other collaborators
obtained the imaginary part of the effective action, but
directly from the integral expression (3) with the help of the Mittag-Leffler
theorem, {\it without using 
his identity} \cite{m2}. Obviously they could not use the wrong
identity because the pole structure of his identity was different
from that of the integrand in (3). This could be another reason why
the experts in this field did not trust the work of \cite{miel}.
Now, curiously the results in \cite{miel} and \cite{m2} combined together
give an expression of the effective action which is almost equivalent
to our result (4), except the logarithmic correction 
in the classical part of the action that we
discuss in the following. How could this have been possible?
We are not in a position to answer this interesting question,
but remark that one could have obtained the results in \cite{miel}
and \cite {m2} simply by integrating the expressions in \cite{clau}.

Now, Jentschura and Weniger in their revised Comment correctly 
remarked that one is not required to obtain the real and imaginary parts
of the effective action simultaneously. We never asserted that 
one is required to do so. Instead
what we emphasized was the fact that {\it if one uses the correct identity,
the integral expression automatically gives us both the real and imaginary
parts simultaneously}. One does not have to re-calculate the imaginary
part with an independent method. Mielniczuk could 
not do this in \cite{miel} with his incorrect
identity, and had to use a different method to obtain the
imaginary part in \cite{m2} eleven years later.

Another important difference is the logarithmic correction 
$\ln (m/ \mu)^2$ of the classical part of the action in our expression (4). 
Obviously one can make this term vanish simply by choosing 
$\mu$ to be $m$. But this is not the point.
The point here is that the effective action in general must be a function
of the subtraction parameter $\mu$ (or equivalently the running scale
after the renormalization). More importantly
the logarithmic correction in our expression tells that
there should exist a finite quantum correction to the classical
Lagrangian $-(a^2 - b^2)/2 e^2$ which is physical.  
Indeed after the renormalization we do obtain \cite{cho2},
\bea
&{\cal L}_{\rm ren} =
-\dfrac{a^2-b^2}{2\bar e^2}{\Big (}1-\dfrac{\bar e^2}{2 \pi^4}
\zeta_1(\bar x){\Big )} \nn\\
&- \dfrac{1}{4\pi^3} ab  \sum_{n=1}^{\infty}\dfrac{1}{n}
{\Big [}\coth(\dfrac{n \pi b}{a}){\Big (} {\rm ci}(\dfrac{n \pi m^2}{a})
\cos(\dfrac{n \pi m^2}{a}) 
 +{\rm si}(\dfrac{n \pi m^2}{a}) \sin(\dfrac{n \pi m^2}{a}){\Big )} \nn\\
&-\dfrac{1}{2} \coth (\dfrac{n \pi a}{b}) {\Big (} \exp(\dfrac{n \pi m^2}{b})
{\rm Ei}(-\dfrac{n \pi m^2}{b}) 
+ \exp(-\dfrac{n \pi m^2}{b}){\rm Ei}(\dfrac{n \pi m^2}{b}
-i\epsilon){\Big )}{\Big ]},
\eea
where $\bar e$ is the running coupling constant which approaches to
$e_{exp}$ asymptotically, 
\bea
&\zeta_1 (\bar x) = {\Big (} f(\bar x)-\bar x \dfrac{df(\bar x)}
{d\bar x}+\dfrac{\bar x^2}{2} 
\dfrac{d^2 f(\bar x)}{d\bar x^2} {\Big )}, \nn\\
&f(\bar x) =\dfrac{}{}\sum_{n=1}^{\infty}\dfrac{1}{n^2}{\Big(}{\rm ci}(n\bar x) 
\cos (n\bar x)
+ {\rm si} (n\bar x) \sin (n\bar x) {\Big)}, 
~~~~~\bar x = \dfrac{\pi m^2}{\bar \mu^2},
\eea
and $\bar \mu$ is the running scale.
This shows that at a finite $\bar x$ we have a finite quantum
correction $\zeta_1 (\bar x)$ to the classical part of the action.

Now, Jentschura and Weniger in their revised Comment insist that 
our logarithmic correction had not been present in our first
version of \cite{cho1}. But the careful readers will find that
the logarithmic term {\it was} present in $I_1$ of (8) in \cite{cho1}.
Furthermore they insist that ``if this correction term is 
of physical significance, all textbook treatments of the problem
would have to be rewritten''. And they quote Schwinger, ``The
logarithmic divergent factor that multiplies the Maxwell lagrangian
may be absorbed by ..... renormalization of charge'', 
to support their claim. Here again they distort the truth. 
First of all, {\it no textbook (as far
as we know) has ever discussed the renormalization of the 
effective action in the present context either with our result or  Mielniczuk's result}.  
We discussed the renormalization of the effective action 
in \cite {cho2} for the first time. So, if there is any reason why 
``all the textbooks have to be rewritten'', it would be because 
the textbooks so far did not discuss the renormalization of the 
effective action (4). Notice that the renormalization of the 
perturbative theory and that of the non-perturbative effective action
produce different results, for example, different 
$\beta$-function \cite{cho2}. This renormalization of the effective action has 
never been discussed in the existing textbooks.

The really relevant
question here is whether the Maxwell term acquires any physical
correction after the renormalization. Our result (6) explicitly 
shows that it does acquire the $\zeta_1 (\bar x)$-correction.
Does this contradict with the Schwinger's remark?  
{\it Obviously Not!} On the contrary it endorses his remark. To see this
notice that $\zeta_1 (\bar x)$ approaches to zero when $\bar x$ goes
to infinity. In this limit the running coupling approaches
to the experimental fine structure constant, so that the quantum correction
to the Maxwell term does disappear with this renormalized coupling constant.
This is exactly what Schwinger intended to explain. 

What we emphasize here is the fact that {\it for a finite $\bar x$,
there is a real quantum correction to the Maxwell term}. 
This is a new observation, which we demonstrated with our 
expression (4) in \cite{cho2}. 

Of course, Mielniczuk has {\it never} discussed any
renormalization whatsoever. Without the renormalization
one can not claim that the coupling constant
in \cite{miel} is the renormalized one.
Under this circumstance  
Jentschura and Weniger should explain why they insist to interpret the coupling
constant in \cite{miel} as the renormalized (i.e., experimental) one. 

\section {Effective Action of Scalar QED and Duality}

Thirdly, our paper contains two more important results, the derivation 
of the convergent series expression of the scalar QED and the establishment
of the electric-magnetic duality in the effective action of QED,
which are as important as our first result.
Curiously, Jentschura and Weniger completely neglected these important results.
We emphasize that {\it the derivation of the effective action of the scalar QED
in terms of a non-perturbative but convergent series expression is 
a non-trivial feat, which is based on a totally new identity
of us} (i.e., the identity (21) of \cite{cho1})
that is as significant as the Sitaramachandrarao's identity. 

As importantly our duality, the invariance 
of the quantum effective action under the transformations
\bea
a \rightarrow -ib,~~~~b \rightarrow  ia,
\eea
is a non-trivial new symmetry. In his new paper
\cite{bug} Jentschura discussed our duality without quoting us.
Furthermore he has trivialized the duality by claiming that 
this duality ``immediately follows
from the integral representation (3).''
But again this is a totally wrong and misleading statement. 
In fact it is easy to see 
that the integral expression (3) itself
is invariant
under the four different transformations
\bea 
a \rightarrow \pm ib, \,\,\, b \rightarrow \pm ia ~~or \mp ia. 
\eea
But among the four symmetries only our duality (8) survives as
the true symmetry of the final quantum effective action.  
So it is obvious that our duality
does {\it not} follow immediately from the integral representation. 

Now, in their revised Comment Jentschura and Weniger insist that 
``this duality (i.e., $E \rightarrow iB,~B \rightarrow -iE$ 
in their language) has been 
known'', {\it without quoting who introduced this duality
first, and  in what context}. It has been known that the
effective action with the pure electric background can be obtained
from the effective action with the pure magnetic background by
the replacement $a$ to $ib$ \cite{ditt}. But our duality
asserts that the effective action with an arbitrary constant
background should be invariant under the transformation (7).
We have established this duality as
a fundamental symmetry of the quantum effective action of QED \cite{cho1,cho3}.
As far as we understand, nobody has proposed the duality in this
context. Obviously they could end this dispute 
(and prove who is telling the truth) simply by
quoting who proposed this duality first under what context. 
We {\it urge} them to do so.

We emphasize that this duality is a 
fundamental symmetry of gauge theories, both Abelian 
and non-Abelian. In fact {\it the duality provides an important
criterion to check the correctness of the quantum effective action
in gauge theories, in particular in QCD} \cite{cho4}.
Jentschura again completely missed this important point. 

\section {Other Issues}

In their revised comment \cite{bugs} Jentschura and Weniger
continue their criticism which has no physical content.
For example, they complain that ``The work \cite{cho1}, even if it had not
appeared in the literature before, ..... does not identify a useful
parameter in terms of which the QED perturbation series would
constitute a convergent series''. But it is well-known that such a
parameter does not exist \cite{dyson}, which was precisely our starting
point in \cite{cho1}. There are other minor criticisms 
(e.g., the complaint on our notation etc.....) which we don't
feel worth responding. We welcome all constructive criticism,
but not the destructive ones which have no physical content.
Here we simply hope that the above response is enough
for the readers to judge the correctness of \cite{bugs}. 

\section {Conclusion}

To sum up, it is evident that the criticism by Jentschura 
and Weniger \cite{bugs} 
on our work \cite{cho1} is a totally unfounded and false accusation.
As we have pointed out it is made of the intentional distortion of truth.
Furthermore Jentschura's new paper \cite{bug} misrepresents and trivializes
our duality without understanding the deep meaning,
without quoting who introduced this duality first.

Finally we would like to emphasize that the purpose of this Reply
is {\it not} to criticize \cite{miel}. We have no intention to do so.
Here we are simply responding to the incorrect Comment of 
\cite{bugs}.


\begin{thebibliography}{99}

\bibitem {cho1} Y. M. Cho and D. G. Pak, hep-th/0006057.
\bibitem {bugs} U. D. Jentschura and E. J. Weniger, hep-th/0007108.
\bibitem {miel} W. Mielniczuk, J. Phys. {\bf A15}, 2905 (1982).
\bibitem
{sita} R. Sitaramachandrarao,  in {\it Ramanujan's Notebooks Vol. 2},
p271, edited by
   B. C. Berndt  (Springer-Verlag) 1989.
\bibitem {m2} S. Valluri, D. Lamm, and W. Mielniczuk, Can. J. Phys.
{\bf 71}, 389 (1993).
\bibitem  {p1} 
S. Blau, M. Visser, and A. Wipf, Int. J. Mod. Phys. {\bf A6}, 5409 
(1991); J. Heyl and L. Hernquist, Phys.
Rev. {\bf D55}, 2449 (1997); R. Soldati and L. Sorbo, Phys. Lett. {\bf B426}, 82 (1998);
G. Dunne and T. Hall, Phys. Rev. {\bf D60}, 065002 (1999); C. Beneventano and E. Santangelo, hep-th/0006123.
\bibitem {p2} W. Dittrich and M. Reuter, {\it Effective Lagrangians in Quantum
Electrodynamics}, Lecture Notes in Physics Vol. 220 (Springer) 1985;
W. Dittrich and H. Gies, {\it Probing the Quantum Vacuum}, Tracts 
in Physics Vol. 166 (Springer) 2000.
\bibitem{clau} M. Claudson, A. Yilditz, and P. Cox, Phys. Rev. {\bf D22},
2022 (1980).
\bibitem {cho2} Y. M. Cho and D. G. Pak, hep-th/0010073.
\bibitem {bug} U. D. Jentschura, physics/0010038.
\bibitem {ditt} See for example, W. Dittrich, W. Tsai, and K. Zimmermann, 
Phys. Rev. {\bf D19}, 2929 (1979).
\bibitem {cho3} W. S. Bae, Y. M. Cho, and D. G. Pak, hep-th/0011196,
\bibitem {cho4} Y. M. Cho and D. G. Pak, hep-th/0006051,
   in {\it Proceedings of TMU-YALE Symposium on Dynamics of Gauge Fields},
edited by
  T. Appelquist and H. Minakata (Universal Academy Press, Tokyo) 1999.
\bibitem
{dyson} F. Dyson, Phys. Rev. {\bf 85}, 631 (1952);
A. Zhitnitsky, Phys. Rev. {\bf D54}, 5148 (1996).
\end{thebibliography}
\end{document}